# Fundamental Analysis of a Developer Support Chat Log for Identifying Process Improvement Opportunities

Zádor Dániel Kelemen

Daniel.Kelemen@NNG.com

**Abstract** In this report analysis of a support chat log of a development team is shown. Developer support chat is used to provide internal support to other development teams. The report shows how a fundamental data analysis helped to identify gaps and action items to boost performance of a development team by reducing time spent on developer support chat and minimizing interrupts from other developer teams. The report also shows an example of how a root cause analysis can be supported by simple data analysis in finding process improvement opportunities.

**Keywords:** metric, measure, KPI, Skype, IT Support, Service, chat, RCA, Root Cause Analysis, co-located teams, interrupt, development time, chat log analysis

## 1  Introduction

Classic management and process improvement frameworks use measurement as a tool for continuous improvement [1]. Upper levels of widespread quality approaches [2] such as CMMI [3] or (Automotive) SPICE [4] recommend quantitative management of work products and processes. These approaches focus on "what" to be done, and thus they intentionally do not provide further guidance on "how" to analyse data or "from where to collect useful data".

This report shows how action items of a process improvement were identified at a core development team supporting a high number of projects (and high number of other developers) based on simple analyses of data extracted from a modern messaging application.

At the beginning of this analysis, management of a business unit at NNG LLC. requested a root cause analysis [5] to investigate why technical support time is high at a core team. NNG is a global leader of automotive navigation software, has 700+ employees in different locations, most of them in 5 different buildings in Budapest, Hungary [6]. After several discussions with the management representatives and interviews with the technical staff it came out that the amount of time spent on developer support chat is unknown. It also came out that one main channel of technical support requests is a Skype[7] chat messages having members from all over the organisation besides the members of the core team. From development time view these are considered interrupts as.

Drawback of developer time interrupts has been investigated by multiple researchers and it was shown that interrupts can have negative effects on software development performance and the cost and effects (e.g. recovery time after an interrupt) can be quantified [8]–[10].

The scope of this report is to focus only on the chat log analysis and on identifying process improvement opportunities based on data analysis - especially on those related to the reduction of support time. It is not in the scope to to describe the whole root cause analysis performed at the core team.

Section 2 describes the approach used in this report, section 3 presents the data collection and data preparation, section 4 shows simple data analysis on collected data and section 5 provides a brief summary of possible improvements based on analysis results. The report ends with limitations in section 6 and conclusions in section 7 respectively.

## 2 Approach

The initial question of the RCA was to investigate why the development team spends a high amount of time on technical support. This question was answered within the RCA (not scope of this report), and a Skype log analysis was performed as a supportive activity to the RCA in order to identify improvement opportunities related to reducing support time. Therefore for the skype log analysis (scope of this report) the question is: How time spent on technical support chat could be reduced based on chat log analysis?

In order to answer the question the following steps were identified:

I. Data collection and data preparation (discussed in 3),

II. Data analysis (discussed in 4),

III. Identify gaps and improvement opportunities based on data analysis results (discussed in 5).

## 3 Data collection and data preparation

In order to perform an organisation-related data analysis of a support chat two types of data are needed: (1) the logs of the chat and the (2) organisational related data such as roles and composition of teams.

The following steps were performed during the data collection and data preparation phase:

1. Data collection from support chat

A 6 months chat log was provided by a team leader involving 140+ active days which was considered sufficient for a fundamental analysis. Data from Skype has been collected by making use of SkypeLogView tool [11].

2. User data collection from internal database

    At NNG, list of employees, teams and various contact information including Skype are stored in a database. Therefore it was needed to collect data from the organisational database and the Skype chat and to merge these based on unique Skype IDs. Since the internal database contains the skype IDs no identity matching algorithms (e.g. such as [12]) were needed. Skype IDs of active users of the support chat were used as a search key in the internal database.

3. Data preparation

    Data preparation consisted of merging user information collected from the internal database with the data collected from support chat log data. All activities were considered as message sending (broadcasting) and end-line characters were removed from multi-line messages (thus considered as one message). Messages sent with the same timestamp by the same user were considered as multi-line messages.

## 4 Data analysis

At the beginning of the analysis measure/metric candidates were identified in a brainstorming: total number of users, number of active users in a period, number of inactive users in a period, total messages, number of messages of the investigated team, external messages, total messages per user (most and less active users), total messages per role (most active role), average number of messages / day, average number of messages / hour and conversation length.

A period of **~6 months** has been analysed (2014.7.7-2015.2.11), including **218** days of which **144** were active days having 3529 messages in total. A day is considered active day when at least one message is sent. Only partial data were available on the first and last days, therefore in some of the analyses these two days were excluded, taking into account only 142 days (e.g. when calculating daily averages) with 3498 messages in total.

Measurements cost/benefit ratio is always a central question. Therefore it was decided that only quick and simple measurements will be performed and not all possible the analyses. For example "conversation length" was excluded because it could be difficult to measure real length of a conversation. Skype chats are working in a broadcasting mode: all members get all messages. Thus, it is difficult to identify attributes of conversations such as start time, end time, interrupt times by another conversation start, all those which are needed to identify conversation lengths.

Technical Report #NNGTR-CTUQM-1501

Taking into the account the metrics cost/benefit ratios the following set of the metrics were identified to be measured: messages per day, hourly distribution of messages, distribution of messages per weekdays, active versus inactive users, activeness of teams, activeness of roles and behaviours of top active users. In this section these metrics are discussed.

## 4.1 Messages per day

In order to identify peak days, message distribution per active days was checked for first.

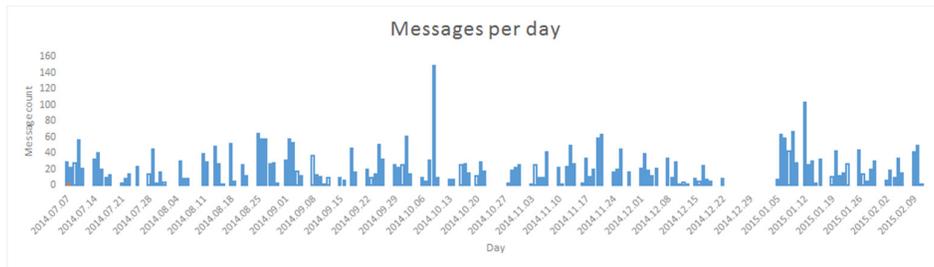

Figure 1 – Distribution of messages per day

Figure 1 shows the message distribution for the entire analysed period while Table 1 lists top 10 peak days (peak – number of messages is highest compared to other days).

Table 1 - Top 10 peak days

| *Date* | *Messages* |
|---|---|
| *2014.10.09* | 149 |
| *2015.01.12* | 103 |
| *2015.01.09* | 67 |
| *2014.08.25* | 64 |
| *2014.11.21* | 63 |
| *2015.01.06* | 63 |
| *2014.10.02* | 61 |
| *2014.11.20* | 59 |
| *2015.01.07* | 59 |
| *2014.08.27* | 58 |

## 4.2 Hourly distribution of messages

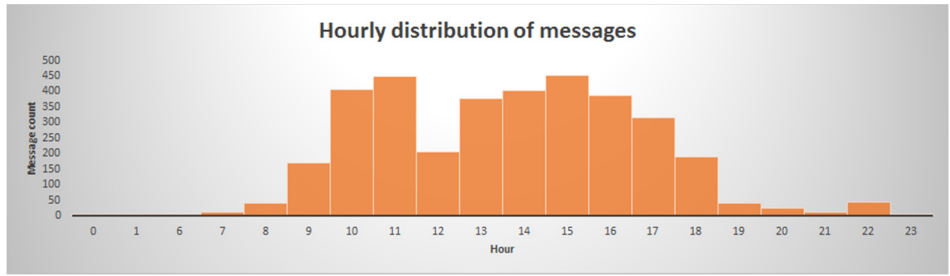

Figure 2 – Hourly distribution of messages

It seemed useful (may be used for future improvements) to investigate peak hours in support time. Figure 2 and Table 2 show the hourly distribution of messages. Peak hours are **15-16** and **11-12**.

Table 2 – Message distribution per hour

| Hour | Messages | Average |
|---|---|---|
| 15 | 452 | **3,18** |
| 11 | 449 | **3,16** |
| 10 | 405 | 2,85 |
| 14 | 402 | 2,83 |
| 16 | 388 | 2,73 |

## 4.3 Peak days

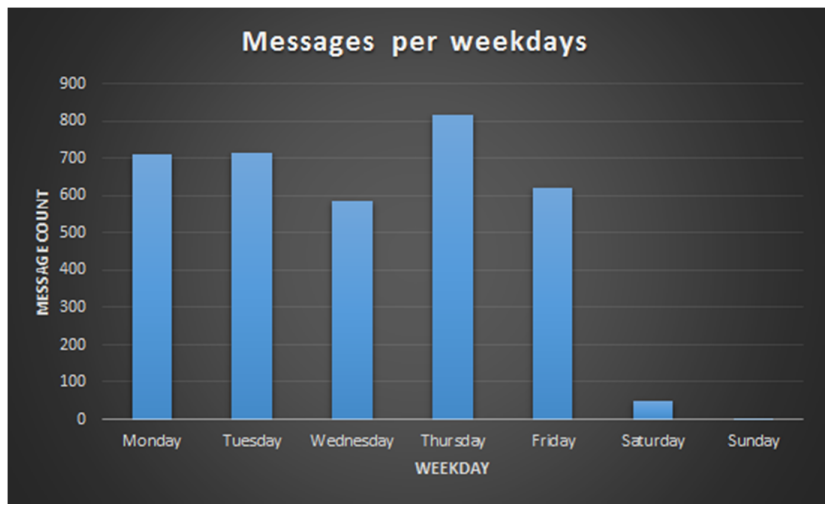

Figure 3 – Distribution of messages per weekday

Figure 3 and Table 6 in the appendix show messages per weekdays. It can be seen that (1) some messages were sent on weekends and (2) there are no peak days.

## 4.4  Active vs inactive users

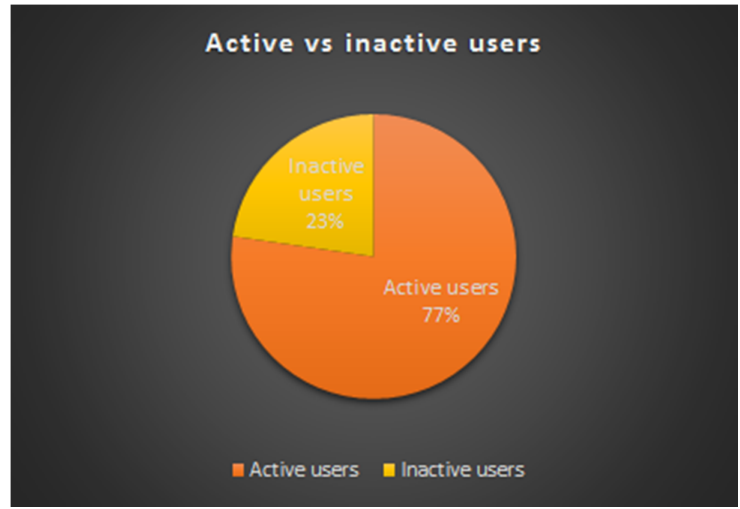

Figure 4 – Active versus inactive users

Since messages are broadcasted, it was interesting to see what percentage of users were inactive in the period analysed. Figure 4 and Table 7 show that 23% of users (35 out of 154) were inactive in the analysed period. NNG has 700+ employees meaning that 22% of employees are member of this support chat of which 17% of all employees were active in the investigated 6 months.

## 4.5  Activeness of teams

Table 3 and Figure 5 show activeness of teams and units. 5 external business unit asked the help of the team. It also can be seen that most of the messages were sent by the investigated team, most of the messages were sent within the business unit.

**Legend for** Table 3**:**
*BU / Team*: Business Unit or Team.
*Messages*: Total messages / Business Unit or Team.
*Active users*: Active users / Business Unit or Team.
*Active user messages in a team*: Total messages of a team / number of active users in a team
*Avg. messages /team /active day*: Average number of messages of a team per active day.
*Avg. messages/active user/day in a team*: Average number of messages / active users / active day / Business Unit or Team
*Active user*: a user who sent messages during the investigated period

*Active day*: a day on which at least one message is sent

Table 3 – Activity of units / teams

| BU / Team | Messages | Active users | Active user messages in a team | Avg messages /team /day | Avg messages/active user/day in a team |
|---|---|---|---|---|---|
| *Average* | 3498 | 119 | 29.39 | 24.63 | 0.21 |
| *External unit 1* | 806 | 23 | 35.04 | 5.68 | 0.25 |
| *External unit 2* | 680 | 30 | 22.67 | 4.79 | 0.16 |
| *External unit 3* | 22 | 2 | 11.00 | 0.15 | 0.08 |
| *External unit 4* | 231 | 17 | 13.59 | 1.63 | 0.10 |
| *External unit 5* | 21 | 3 | 7.00 | 0.15 | 0.05 |
| *Other* | 34 | 4 | 8.50 | 0.24 | 0.06 |
| *Unit of the investigated team* | 1704 | 40 | 42.60 | 12.00 | 0.30 |
| *Parent team of investigated team* | 1151 | 14 | 82.21 | **8.11** | **0.58** |
| *Investigated team* | 979 | 9 | **108.78** | 6.89 | 0.77 |

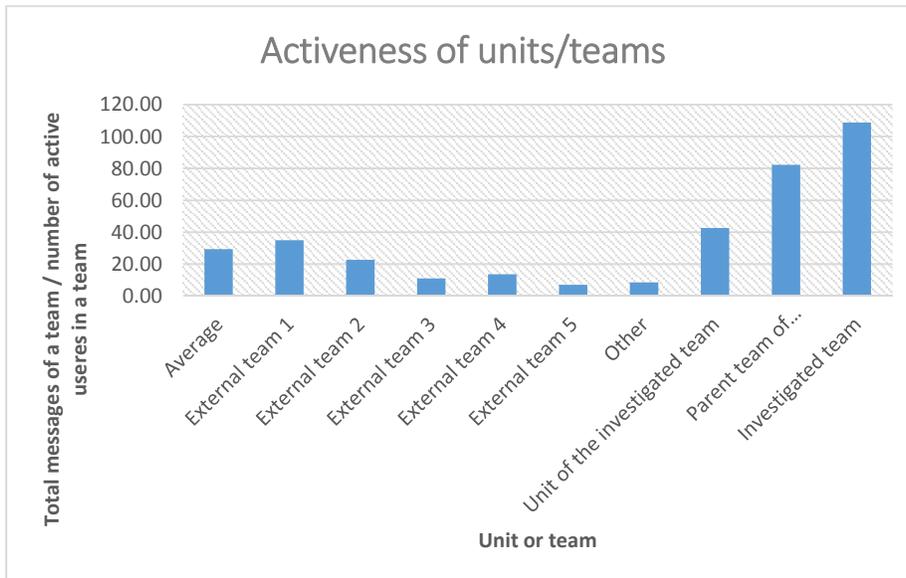

Figure 5 – Activeness of teams

Technical Report #NNGTR-CTUQM-1501

## 4.6 Activeness of roles

It was interesting to see if the chat is filling its purpose, to see who (in what role) are sending messages on the chat. It came out that the majority of messages (68%) are sent by developers (see Figure 6).

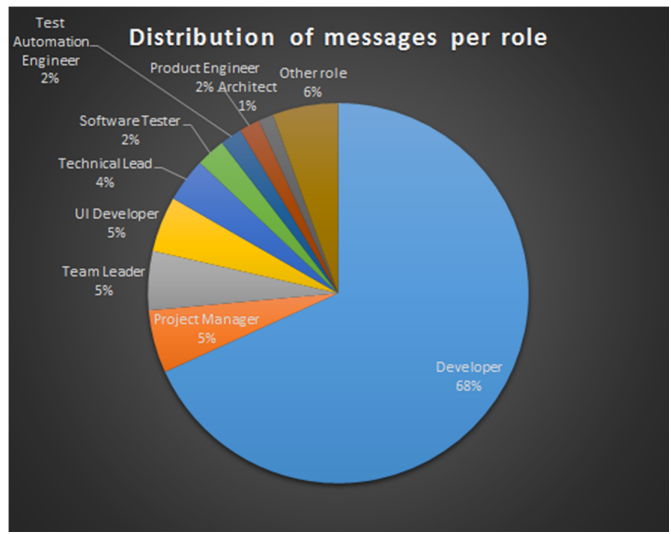

Figure 6 – Distribution of messages per roles

## 4.7 Top 10 active users

Table 1 and Figure 7 show activeness of top 10 most active users based on message count. It can be seen that the most active user sent 315 messages in total which results in only 2,22 messages per day. It also can be seen that top 4 active users are members of the investigated team.

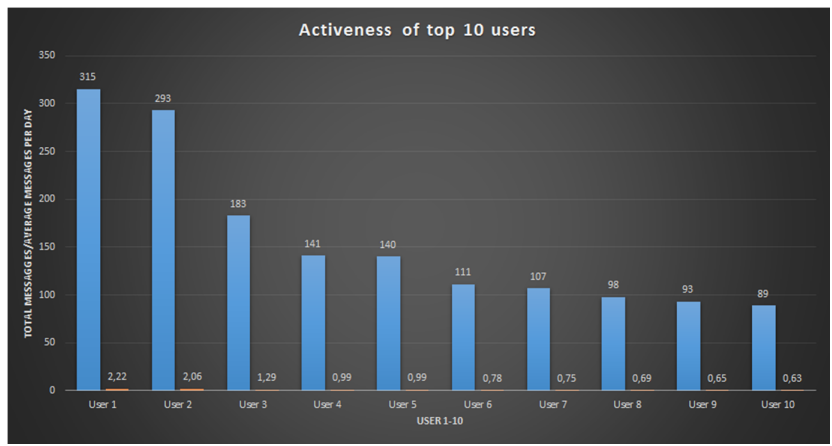

Figure 7 – Top 10 active users

Table 4 – List of top 10 active users

| User Alias | Total messages | Avg message / day | Team | Role |
|---|---|---|---|---|
| *User 1* | 315 | 2,22 | Investigated team | Developer |
| *User 2* | 293 | 2,06 | Investigated team | Developer |
| *User 3* | 183 | 1,29 | Investigated team | Developer |
| *User 4* | 141 | 0,99 | Investigated team | Developer |
| *User 5* | 140 | 0,99 | External team | Developer |
| *User 6* | 111 | 0,78 | External team | Team leader |
| *User 7* | 107 | 0,75 | External team | Developer |
| *User 8* | 98 | 0,69 | External team | Developer |
| *User 9* | 93 | 0,65 | External team | Developer |
| *User 10* | 89 | 0,63 | External team | Developer |

## 5  Improvement opportunities identified based on data analysis

Table 5 shows ID and referred data analysis section (column 1), results deducted from Skype chat log analysis (column 2) and gaps identified (column 3).

Table 5 – Analysis results and gaps

| ID (section) | Analysis result | Gap |
|---|---|---|
| *S-F1* (4.1) | There are days on which support chat interaction is high. In these days it is common that 20+ messages are sent in peak hours resulting in an interrupt in every ~3 minutes. In case if users are listening, their day is practically lost on peak support days. | Developers are not protected from interrupts |
| *S-F2* (4.1, 4.3) | It is unexpected when a peak support day occurs (no trend can be derived) | Developers are not protected from interrupts |
| *S-F3* (4.2) | Peak support hours overlap peak developer hours (core office hours are between 10-16) | Developers are not protected from interrupts |
| *S-F4* (4.6) | 68% of interaction is by developers | No support role exist for support tasks |

| ID (section) | Analysis result | Gap |
|---|---|---|
| *S-F5 (observed, not in 4)* | It is difficult to search in (Skype) support chat log and new users have no access to the skype log, same questions may happen in future. | There is no knowledge base |
| *S-F6 (4.5, 4.6, 4.7)* | Top commenters send 0,6-2,2 messages per day and the investigated team members send 0,7 messages per day in average, there is no continuous need for all developers to listen the support chat | Developers are not protected from interrupts (while they could be protected!) |
| *S-F7 (4.4)* | There are 154 users of the support chat. 23% of users were inactive in the last 6 months. Many of them may be interrupted, especially during peak support days (they may delete or mute support chat to avoid interrupts) | No support chat mute guide |

Analysis results and gaps served as an input to the investigated team and to the quality management to identify process improvement opportunities (action items with responsibles and deadlines). Not all of them can be listed within the frame of this report due to confidentiality reasons. However, some of them which can be shared publicly were: protect developers from interrupts by (1) define a dispatcher service policy (2) with a weekly rotating dispatcher role, (3) develop and maintain a knowledge base and FAQ page to reduce the number of interrupts of the dispatcher and developers and (4) define and institutionalize a support chat mute guide for inactive users (with chat message keywords for activation).

## 6 Limitations

**Input data** – only a half year log of a team was used, however further is needed to be to analysed in order to gain a more holistic view and to possibly refactor the support activities of other teams.

**In-depth analysis, selection of metrics to be analysed** - obviously there is room to define further, more complex metrics and to use more appropriate views for the analysis.

**Inactive days and inactive users** – only active days (on which data was sent) has been used and only active users were included into the analysis. Analysing all chat members could provide more accurate results, however gaps identified are expected to be similar.

**Changes in roles and within organisation** – a mid-size IT organisation, especially if it is transforming from a start-up to a multinational company has many changes even within a half-year period. These changes were not taken into account (e.g. there were multiple changes within the investigated team: role changes or changes among teams).

# 7 Conclusion

This report showed that root cause analysis can be combined with other techniques such as fundamental data analysis in order to obtain more insights on the attributes of an investigated problem. The scope of this report was to answer the question: "How time spent on technical support chat could be reduced based on chat log analysis?"

In order to answer the question, 7 metrics were analysed and 7 conclusions were deducted, which helped to identify 4 gaps serving the basis for identifying 4 action items. The analysis showed that developers are interrupted many times during core developer hours by support chat requests. A conclusion was made based on data analysis that support chat interrupts at the investigated team can easily be reduced and developers can be protected by implementing action items identified in section 5, namely: (1) definition of a dispatcher service policy with a (2) weekly rotating dispatcher role, (3) development and maintenance of a knowledge base and (4) definition and institutionalization of a support chat mute guide.

At the time of writing of this report the team has already implemented some of the action items: dispatcher role is defined by the team leader, dispatching ideas were collected and summarized in a dispatcher policy - co-authored by the team and externals. When forming the dispatching service policy ITIL [13] and the advantages of T-shaped people [14] were also taken into account. The first dispatcher weeks have already ended.

Efficiency of the new settings will be measured and action items will be tracked until closure by the quality management together with the team leaders.

As a future direction, further techniques (e.g. data/text/process mining approaches, performance evaluation, formal methods, probability theory, polling systems or complexity analysis [15], [16]) and tools (such as ProM, ProM for RapidMiner or DISCO) could be involved to conduct a more detailed data analysis with the goal of understand underlying processes and analysing behaviour of support chats.

Interrupts are not (and probably cannot be) fully eliminated as new problems arise with new developments. As an additional future direction, the introduction of interrupt recovery techniques [17], [18] could also be investigated.

# Acknowledgement

I would like to thank Nicola Grapputo for management support and Sándor Hodosi and Balázs Tódor for useful thoughts and comments.

## About the author

Zádor Dániel Kelemen wrote his first on-the-fly chat robots and log analysers back in the IRC era. His BSc thesis focused on a chat assisted e-learning system, winning the 3rd prize on the scientific contest Hungarian students of Transylvania. After moving his focus to process improvement, he got his MSc degree from Budapest University of Technology and Economics, Hungary and his PhD degree from Eindhoven University of Technology, Netherlands, both in the field of Software Process Improvement. He worked in the field of software process improvement and quality assurance at SQI and ThyssenKrupp Presta. He is currently the Quality Manager of Common Technology Business Unit at NNG Ltd, Hungary.

# Appendix

Table 6 - Distribution of messages per weekday

| Action Time | Weekday |
|---|---|
| Monday | 709 |
| Tuesday | 713 |
| Wednesday | 586 |
| Thursday | 817 |
| Friday | 621 |
| Saturday | 49 |
| Sunday | 3 |
| Grand Count | 3498 (first and last day excluded) |

Table 7 – Active versus inactive support chat members

| | Number of participants | Percent |
|---|---|---|
| All members | 154 | 100 |
| Active members | 119 | 77 |
| Inactive members | 35 | 23 |

Table 8 – Message distribution per role

| Role | Total messages | Percent |
|---|---|---|
| Developer | 2386 | 68,21 |
| Project Manager | 187 | 5,35 |
| Team Leader | 176 | 5,03 |
| UI Developer | 165 | 4,72 |
| Technical Lead | 131 | 3,74 |
| Software Tester | 87 | 2,49 |
| Test Automation Engineer | 65 | 1,86 |
| Product Engineer | 61 | 1,74 |
| Architect | 45 | 1,29 |
| Other role | 195 | 5,57 |